# BOOTSTRAP METHODS IN ECONOMETRICS

by

Joel L. Horowitz
Department of Economics
Northwestern University
Evanston, IL 60208 U.S.A.
joel-horowitz@northwestern.edu

## ABSTRACT

The bootstrap is a method for estimating the distribution of an estimator or test statistic by resampling one's data or a model estimated from the data. Under conditions that hold in a wide variety of econometric applications, the bootstrap provides approximations to distributions of statistics, coverage probabilities of confidence intervals, and rejection probabilities of hypothesis tests that are more accurate than the approximations of first-order asymptotic distribution theory. The reductions in the differences between true and nominal coverage or rejection probabilities can be very large. In addition, the bootstrap provides a way to carry out inference in certain settings where obtaining analytic distributional approximations is difficult or impossible. This article explains the usefulness and limitations of the bootstrap in contexts of interest in econometrics. The presentation is informal and expository. It provides an intuitive understanding of how the bootstrap works. Mathematical details are available in references that are cited.

# 1. INTRODUCTION

The bootstrap is a method for estimating the distribution of an estimator or test statistic by resampling one's data or a model estimated from the data. It amounts to treating the data as if they were the population for the purpose of evaluating the distribution of interest. Under mild regularity conditions, the bootstrap yields an approximation to the distribution of an estimator or test statistic that is at least as accurate as and often more accurate than the approximation obtained from first-order asymptotic theory. Thus, the bootstrap provides a way to substitute computation for mathematical analysis if calculating the asymptotic distribution of an estimator or statistic is difficult, and it often provides a practical way to improve upon first-order approximations.

Improvements on first-order approximations are called asymptotic refinements. The bootstrap's ability to provide asymptotic refinements is important, because first-order asymptotic theory often gives poor approximations to the distributions of test statistics with the sample sizes available in applications. Therefore, the nominal probability that a test based on an asymptotic critical value rejects a true null hypothesis or the nominal coverage probability of a confidence interval can be very different from the true rejection probability (RP) or coverage probability. The information matrix test of White (1982) is a well-known example of a test in which large finite-sample errors in the RP can occur when asymptotic critical values are used (Horowitz 1994, Kennan and Neumann 1988, Orme 1990, Taylor 1987). The bootstrap often provides a tractable way to reduce or eliminate finite-sample errors in the RP's of statistical tests.

The bootstrap was introduced by Efron (1979) and was the object of much research in the ensuing 25 years. Most of the theory of the bootstrap and methods for implementing it were developed during this period. The results of this research are synthesized in books by Beran and Ducharme (1991), Davison and Hinkley (1997), Efron and Tibshirani (1993), Hall (1992),

Mammen (1992), and Shao and Tu (1995). Hall (1994), Horowitz (1997, 2001), Jeong and Maddala (1993), and Vinod (1993) provide reviews with an econometric orientation. In addition to topics in these books and reviews, this article summarizes several recent extensions of bootstrap methods. These include methods for obtaining confidence bands in nonparametric estimation, methods for carrying out non-asymptotic inference in parametric models, and methods for high-dimensional models.

This article explains and illustrates the usefulness and limitations of the bootstrap in contexts of interest in econometrics. The presentation is informal and expository. Its aim is to provide an intuitive understanding of how the bootstrap works and a feeling for its practical value in econometrics. It does not provide a mathematically detailed or rigorous treatment of the theory of the bootstrap. Mathematically rigorous treatments are available in the books by Beran and Ducharme (1991) and Hall (1992) and in references that are cited later in this article.

Although the bootstrap is often very accurate, it can be inaccurate and misleading if it is used incorrectly. Examples include inference about a parameter that is on the boundary of the parameter set, inference about the maximum or minimum of random variables, and inference in the presence of weak instruments. The difference between correct and incorrect use is not always intuitive. A further objective of this article is to explain the difference between correct and incorrect use

The variety of different bootstrap methods and applications and the associated literature are very large. A given bootstrap method typically works well in some settings but not in others. It is not possible to provide an exhaustive review of bootstrap methods or the bootstrap literature in one article, and this article does not attempt to do so. Instead, the article focusses on a few methods and settings that arise frequently in applied econometrics. Owing to space limitations, this article does not present Monte Carlo or other illustrations of the numerical performance of the bootstrap.

Such illustrations are provided in Horowitz (2001) and other references cited throughout the article.

The remainder of this article is divided into eight sections. Section 2 explains the bootstrap sampling procedure and gives conditions under which the bootstrap distribution of a statistic is a consistent estimator of the statistic's asymptotic distribution. Section 3 explains when and why the bootstrap provides asymptotic refinements. This section concentrates on data that are independent random samples from a distribution and statistics that are either smooth functions of sample moments or can be approximated with asymptotically negligible error by such functions (the smooth function model). Section 4 extends the results of Section 3 to statistics that do not satisfy the assumptions of the smooth function model. Section 5 describes the use of the bootstrap in nonparametric estimation of a conditional mean or quantile function. Section 6 describes the multiplier and wild bootstraps. Section 7 presents results on the application of the bootstrap to statistics based on LASSO and adaptive LASSO parameter estimates. Section 8 discusses the application of the bootstrap to dependent data, and Section 9 presents concluding comments.

## 2. THE BOOTSTRAP WITH INDEPENDENTLY AND IDENTICALLY DISTRIBUTED DATA

Much research in empirical economics is based on data that are a random sample from a distribution. This type of data includes panel-data based on randomly sampled individuals, where the data consist of observations of individuals and other relevant variables over time. To minimize the complexity of the discussion in this section, let the data consist of a random sample of size $n$ from the distribution of a possibly vector valued random variable $X$. If the data are used in a model that has dependent and explanatory variables, both types of variables are included among the components of $X$. Denote the data by $\{X_i : i = 1,...,n\}$. Let $F_0$ denote the distribution function of $X$ and $F_n$ denote the empirical distribution function of the sample. That is, for any vector $z$ whose dimension is the same as that of $X$,

$$F_n(z) = \frac{1}{n} \sum_{i=1}^{n} I(X_i \leq z),$$

where $I(\cdot)$ is the indicator function and the inequality holds component-by-component. Suppose that the object of interest is a statistic $T_n(X_1,...X_n)$. For example, if $X$ is a scalar, $T_n$ might be the $t$-statistic for testing the null hypothesis $H_0: E(X) = 0$, where $E$ is the expectation operator and $E(X)$ is the population mean of $X$. In this case,

$$T_n(X_1,...,X_n) = \overline{X} / s_n,$$

where $\overline{X}$ is the sample average and $s_n$ is the sample standard deviation:

$$\overline{X} = \frac{1}{n} \sum_{i=1}^{n} X_i; \qquad s_n^2 = \frac{1}{n-1} \sum_{i=1}^{n} (X_i - \overline{X})^2.$$

In most applications, the distribution of $X$ and, therefore, of $T_n$ is unknown. It is necessary to estimate the distribution of $T_n$ to obtain a critical value for testing $H_0$. Asymptotic distribution theory provides one way of estimating this distribution. In this simple example, $T_n$ is asymptotically distributed as $N(0,1)$. Moreover, the $N(0,1)$ distribution provides a very accurate approximation to the unknown exact finite sample distribution of $T_n$ with samples of the sizes typically encountered in empirical economics. In more complicated settings, however, the asymptotic approximation may be inaccurate or the asymptotic distribution of $T_n$ hard to compute.

*2.1 The Nonparametric Bootstrap and Bootstrap Consistency*

The nonparametric bootstrap estimates the distribution of $T_n$ by treating the empirical distribution of the data as if it were the population distribution. In other words, the nonparametric bootstrap conditions on the data pretends that they were drawn from a population whose distribution function is $F_n$ instead of $F_0$. In this pretend world, $X$ has a known distribution (the empirical distribution of the data), and the exact finite-sample distribution of $T_n$ can be computed with any desired accuracy by simulation. The steps of this procedure are:

1. Generate a bootstrap sample $\{X_i^* : i = 1,...,n\}$ by sampling the original data randomly with replacement.

2. Compute $T_n^* = T_n(X_1^*,...,X_n^*)$.

3. Use the results of many repetitions of steps 1 and 2 to compute the empirical probability of the event $T_n^* \leq \tau$ for any $\tau$. This probability is the proportion of repetitions in which the event $T_n^* \leq \tau$ occurs.

Brown (2000) and Hall (1992, Appendix II) discuss simulation methods that take advantage of techniques for reducing Monte Carlo sampling variation. Regardless of the simulation method that is used, however, the essential characteristic of the nonparametric bootstrap is treating $F_n$ as if it, instead of $F_0$, were the distribution function of the population from which the data were sampled.

Now let $G_n(\cdot, F)$ be the distribution function of $T_n$ when the data are sampled randomly from a population whose distribution function is $F$. Thus, the true but unknown distribution function of $T_n$ is $P(T_n \leq \tau) = G_n(\tau, F_0)$. The bootstrap approximates $G_n(\tau, F_0)$ by $G_n(\tau, F_n)$. The Glivenko-Cantelli theorem ensures that $|F_n(z) - F_0(z)| \to 0$ almost surely uniformly over $z$. Therefore, intuition suggests that $G_n(\tau, F_n)$ should be close to $G_n(\tau, F_0)$ if $n$ is large and $G_n$ is continuous at $F_0$ in the sense that $G_n(\tau, F)$ is close to $G_n(\tau, F_0)$ whenever $F$ is close to $F_0$. Let $G_\infty(\tau, F_0)$ denote the asymptotic distribution function of $T_n$. Since $G_n(\tau, F_0)$ converges to $G_\infty(\tau, F_0)$ as $n \to \infty$ by definition, one can hope that $G_n(\tau, F_n) \to G_\infty(\tau, F_0)$ as $n \to \infty$. In other words, the bootstrap estimate of the distribution function of $T_n$ converges to the asymptotic distribution function of $T_n$ as $n \to \infty$. This property of the bootstrap is called consistency. The formal definition of bootstrap consistency is

**Definition**: *Let $P_n$ denote the probability distribution of the sample $\{X_i: i = 1, \ldots, n\}$. The bootstrap estimator $G_n(\cdot, F_n)$ is consistent if for each $\varepsilon > 0$ and $F_0$*

$$\lim_{n \to \infty} P_n\left[\sup_\tau |G_n(\tau, F_n) - G_\infty(\tau, F_0)| > \varepsilon\right] = 0.$$

Beran and Ducharme (1991) give general conditions under which the bootstrap is consistent, but these conditions are hard to check. Mammen (1992) gives less general but much more useful conditions for bootstrap consistency. Specifically, Mammen (1992) gives necessary and sufficient conditions for the bootstrap to consistently estimate the distribution of a linear functional of $F_0$. These conditions are important because they are often easy to check, and many econometric estimators and test statistics are asymptotically equivalent to linear functionals of some $F_0$. Gill (1989) and Hall (1990) give related theorems. Mammen's (1992) theorem is:

**Theorem 2.1** (Mammen 1992): *Let $\{X_i : i = 1, \ldots, n|$ be a random sample from a population. For a sequence of functions $g_n$ and sequences of numbers $t_n$ and $\sigma_n$, define $\bar{g}_n = n^{-1} \sum_{i=1}^{n} g_n(X_i)$ and $T_n = (\bar{g}_n - t_n)/\sigma_n$. For the bootstrap sample $\{X_i^* : i = 1, \ldots, n\}$, define $\bar{g}_n^* = n^{-1} \sum_{i=1}^{n} g_n(X_i^*)$ and $T_n^* = (\bar{g}_n^* - \bar{g}_n)/\sigma_n$. Let $G_n(\tau) = P(T_n \leq \tau)$ and $G_n^*(T_n^*) = P_n^*(T_n^* \leq \tau)$, where $P_n^*$ is the probability distribution induced by bootstrap sampling* *conditional on the original data* $\{X_i\}$. *Then* $G_n^*(\cdot)$ *consistently estimates* $G_n(\cdot)$ *if and only if*

$T_n \to^d N(0,1)$.

Bootstrap sampling is conditional on the original data $\{X_i\}$. Therefore, in bootstrap sampling and estimation, the original data and any functions of the data are non-stochastic. Only functions of the bootstrap sample $\{X_i^*\}$ are random in bootstrap sampling.

Maximum likelihood estimators, generalized method of moments estimators, and other commonly encountered extremum estimators and test statistics are asymptotically linear and asymptotically normal (or are asymptotically chi-square quadratic forms of asymptotic normal statistics) under the usual regularity conditions (e.g., Amemiya, 1985, Ch. 4). Therefore, the nonparametric bootstrap estimates their asymptotic distributions consistently and can be applied to them. Indeed, the bootstrap is often more accurate than asymptotic distribution theory, as will be discussed in Section 3 of this article.

There are, however, important settings in which a statistic is neither asymptotically linear nor normal and for which the nonparametric bootstrap is inconsistent. These settings include:

1. Manski's (1975, 1985) maximum score estimator for a binary response model. In this model, the binary dependent variable $Y$, vector of explanatory variables $X$, and vector of constant coefficients $\beta$ are related by

$$Y = \begin{cases} 1 \text{ if } \beta'X - U \geq 0 \\ -1 \text{ if } \beta'X - U < 0 \end{cases}$$

$P(U \leq 0 | X) = 0.5$.

Let $\{Y_i, X_i : i = 1,...,n\}$ be a random sample from the distribution of $(Y, X)$ and $\|\cdot\|$ denote the $\ell_2$ norm. For any scalar $v$, define

$$\text{sgn}(v) = \begin{cases} 1 \text{ if } v \geq 0 \\ -1 \text{ if } v < 0. \end{cases}$$

The maximum score estimator is

$$\hat{b} = \arg\max_{\|b\|=1} \sum_{i=1}^{n} Y_i \,\text{sgn}(b'X_i)$$

Manski (1975, 1985) gave conditions under which $\hat{b}$ estimates $\beta$ up to an unidentified scale parameter, but the asymptotic distribution of a $\hat{b}$ after suitable centering and scaling is that of the maximum of a multidimensional Gaussian process with quadratic drift (Cavanagh 1987, Kim and Pollard 1990). The nonparametric bootstrap does not estimate this distribution consistently (Abrevaya and Huang 2005).

2. Estimation when a parameter is on the boundary of the parameter set. This situation arises in moment inequality models and estimation under shape restrictions, among other settings, where it is not known whether one or more inequality constraints is binding. The following simple example illustrates the problem. Let $\{X_i : i = 1,...,n\}$ be a random sample from the $N(\mu,1)$ distribution, and suppose it is known that $\mu \geq 0$. Let $\bar{X} = n^{-1}\sum_{i=1}^{n} X_i$ be the sample average of $X$. The maximum likelihood estimate of $\mu$ is $m = \max(0, \bar{X})$. The exact finite-sample distribution of $m$ is given by

$$P[n^{1/2}(m - \mu) \leq z] = \begin{cases} \Phi(z) \text{ if } m \geq 0 \\ 0 \text{ if } m < 0, \end{cases}$$

where $\Phi$ is the standard normal distribution function. This is a censored normal distribution with censoring at the boundary point $m = 0$. If $\mu > 0$, then the probability of censoring approaches 0 as $n \to \infty$, and the limiting distribution of $n^{1/2}(m-\mu)$ is $N(0,1)$. However, if $\mu = 0$, the limiting distribution is

$$P[n^{1/2}m \leq z] = \begin{cases} \Phi(z) \text{ if } z \geq 0 \\ 0 \text{ if } z < 0. \end{cases}$$

It follows from Theorem 2.1 that the nonparametric bootstrap estimates the distribution of $n^{1/2}(m-\mu)$ consistently if $\mu > 0$. However, asymptotic normality of $n^{1/2}(m-\mu)$ is necessary for consistency according to Theorem 2.1. Therefore, the nonparametric bootstrap does not estimate the distribution of $n^{1/2}(m-\mu)$ consistently if $\mu = 0$. Andrews (2000) describes several alternative resampling methods for estimating the distribution of $n^{1/2}(m-\mu)$ consistently when $\mu = 0$. Andrews and Barwick (2012); Andrews and Han (2009); Bugni (2010, 2016); and Bugni, Canay, and Shi (2015, 2017) discuss bootstrap methods for moment inequality models.

3. Distribution of the maximum of a sample. The following example, which is due to Bickel and Freedman (1981), is a simplified version of a situation that occurs in empirical models of auctions and search models, among others. Bowlus, Neumann, and Kiefer (2001); Donald and Paarsch (1996); Flinn and Heckman (1982); and Heckman, Smith, and Clements (1997) describe such models. Chernozhukov and Hong (2004) and Hirano and Porter (2003) discuss some of the theoretical issues that are involved.

Let $\{X_i : i = 1,...,n\}$ be a random sample from the uniform distribution on $[0, \theta_0]$, where $\theta_0$ is a positive constant whose true but unknown value is $\theta_0 = 1$. The maximum likelihood estimator of $\theta_0$ is $\hat{\theta}_n = \max(X_1,...,X_n)$. Define $T_n = n(\hat{\theta}_n - 1)$. As $n \to \infty$, $P(T_n \leq -z) = e^{-z}$ for any $z \geq 0$. Moreover, $P(T_n = 0) = 0$ for all $n$. Let $\{X_i^* : i = 1,...,n\}$ be a bootstrap sample that is obtained by sampling the data $\{X_i : i = 1,...,n\}$ randomly with replacement. The bootstrap analog of $\theta_0$ is $\hat{\theta}_n$, because the bootstrap samples the empirical distribution of the data and $\hat{\theta}_n$ is the upper bound of the support of this distribution. The bootstrap estimator of $\hat{\theta}_n$ is $\theta_n^* = \max(X_1^*,....X_n^*)$. The bootstrap analog of $T_n$ is $T_n^* = n(\theta_n^* - \hat{\theta}_n)$. Let $P_n^*$ denote probability under bootstrap sampling conditional on the original data $\{X_i\}$. Then $P_n^*(T_n^* = 0) = 1 - (1 - n^{-1})^n \to 1 - e^{-1}$ as $n \to \infty$. Thus, the nonparametric bootstrap does not estimate the distribution of $T_n$ consistently.

*2.2 The Parametric and Residual Bootstraps*

Sometimes the distribution of the random variable in a model is assumed to be known up to a finite-dimensional parameter. This happens, for example, in maximum likelihood estimation and in ordinary least squares estimation of a normal linear model with a fixed design. Let $U$ be the random variable or vector in a model and let $F(u,\theta) = P(U \leq u)$ be the cumulative distribution function of $U$, where $F$ is a known function and $\theta$ is a finite-dimensional parameter whose true but unknown value is $\theta_0$. Let $\hat{\theta}_n$ be a consistent estimator of $\theta_0$. If $F(u,\theta)$ is a continuous function of $\theta$ in a neighborhood of $\theta_0$ for all $u$, then bootstrap samples can be drawn randomly from the distribution whose cumulative distribution function is $F(u, \hat{\theta}_n)$.

If the distribution of $U$ is not known, it may be possible to draw bootstrap samples from a model's residuals. As an example, consider the linear model

(2.1)  $Y_i = \theta_0' X_i + U_i; \quad i = 1,...,n$

where the $X_i$'s are fixed in repeated samples and the $U_i$'s are independently and identically distributed random variables with means of zero and finite variances. Let $\hat{\theta}_n$ be the ordinary least squares (OLS) estimator of $\theta_0$, and let $\hat{U}_i = Y_i - \hat{\theta}_n X_i$ ($i = 1,...,n$) denote the OLS residuals. Bootstrap analogs $Y_i^*$ of the $Y_i$'s can be generated from the equation

$$Y_i^* = \hat{\theta}_n' X_i + U_i^*; \quad i = 1,...,n,$$

where the $U_i^*$'s are sampled randomly with replacement from the $\hat{U}_i$'s. This is called the residual bootstrap because it samples the residuals of the estimated model, not the original data. The residual bootstrap sample is $\{Y_i^*, X_i : i = 1,...,n\}$. The bootstrap analog of $\theta_0$ is $\hat{\theta}_n$ and is non-stochastic in bootstrap sampling, though it is random in sampling from the original population. Let $\theta_n^*$ be the OLS estimator of $\hat{\theta}_n$ based on the bootstrap sample. Then the bootstrap distribution of $n^{1/2}(\theta_n^* - \hat{\theta}_n)$ estimates the population distribution of $n^{1/2}(\hat{\theta}_n - \theta_0)$ consistently. Specifically,

$$\lim_{n \to \infty} \sup_{-\infty < z < \infty} |P_n^*[n^{1/2}(\theta_n^* - \hat{\theta}_n) \leq z] - P[n^{1/2}(\hat{\theta} - \theta_0) \leq z]| =^{a.s} 0,$$

where $P_n^*$ is the probability induced by bootstrap sampling of the $U_i^*$'s conditional on the original data $\{Y_i, X_i\}$.

The residual approach can also be used in nonparametric mean and quantile regression. In a nonparametric mean regression, for example, $\theta_0' X_i$ in (2.1) is replaced by $g(X_i) = E(Y | X_i)$, and $\hat{\theta}_n' X_i$ is replaced by a nonparametric estimate of $g(X_i)$. Denote this estimate by $\hat{g}(X_i)$. Denote the residuals of the nonparametric model by $\tilde{U}_i = Y_i - \hat{g}(X_i)$, and let $\hat{U}_i = \tilde{U}_i - n^{-1} \sum_{j=1}^n \tilde{U}_j$ denote residuals that are centered to have a sample average of zero. Residual bootstrap samples $\{Y_i^*, X_i : i = 1,...,n\}$ are generated by setting

$$Y_i^* = \hat{g}(X_i) + U_i^*,$$

where the $U_i^*$'s are sampled randomly from the $\hat{U}_i$'s.

*2.3 Subsampling*

The nonparametric bootstrap draws samples from the empirical distribution of the data, not the population distribution. The nonparametric bootstrap is inconsistent in certain settings because the two distributions are not the same. Although the empirical distribution of the data converges to the population distribution almost surely as the sample size increases, the distribution of the bootstrap analog of a statistic may not converge to the population distribution of the statistic. This problem can be avoided by sampling the true population distribution instead of the empirical distribution of the data. It is not possible to draw repeated samples of size $n$ from the population distribution, but it is possible to draw repeated samples of size $m < n$. This is done by drawing samples of size $m$ randomly without replacement from the estimation data. Each such sample is a subsample of the original data and is a random sample of size $m$ from the population distribution of the data.

This method of subsampling is called non-replacement subsampling. It was originally proposed by Politis and Romano (1994), who show that it consistently estimates the distribution of a statistic under conditions that are much weaker than those required for consistency of the bootstrap estimator. Politis, Romano, and Wolf (1997) extend the non-replacement subsampling method to heteroskedastic time series. Politis, Romano, and Wolf (1999) provide a detailed description of non-replacement subsampling and examples of its application. Bertail, Politis, and Romano (1999) provide a data-based method for choosing a tuning parameter that occurs in non-replacement subsampling.

To describe the non-replacement subsampling method, let $t_n = t_n(X_1,...,X_n)$ be an estimator of the population parameter $\theta$, and set $T_n = \rho_n(t_n - \theta)$, where $\rho_n$ is a normalizing factor that is chosen so that $G_n(\tau, F_0) = P(T_n \leq \tau)$ converges to a nondegenerate limit $G_\infty(\tau, F_0)$ at points $\tau$ where the latter function is continuous. For example, if $\theta$ is a population mean, $t_n = \bar{X}$ (the sample average), and $\rho_n = n^{1/2}$. Let $\{X_{i_j} : j = 1,...,m\}$ be a subset of $m < n$ observations taken from the original sample $\{X_i : i = 1,...,n\}$. Define $N_{nm} = \binom{n}{m}$ to be the total number of subsets of size $m$ that can be formed. Let $t_{mk}$ denote the estimator $t_m$ evaluated at the $k$'th subset. The non-replacement subsampling method estimates $G_n(\tau, F_0)$ by

$$G_{nm}(\tau) = N_{nm}^{-1} \sum_{k=1}^{N_{nm}} I[\rho_m(t_{mk} - t_n) \leq \tau].$$

The intuition behind this method is as follows. Each subsample $\{X_{i_j} : j = 1,...,m\}$ is a random sample of size $m$ from the population distribution whose cumulative distribution function is $F_0$. Therefore, $G_m(\tau, F_0)$ is the exact sampling distribution of $\rho_m(t_m - \theta)$. Moreover,

(2.2) $G_m(\tau, F_0) = EI[\rho_m(t_m - \theta) \leq \tau]$.

The quantity on the right-hand side of (2.2) cannot be calculated in an application because $F_0$ and $\theta$ are unknown. $G_{mn}(\tau)$ is the estimator of $G_m(\tau, F_0)$ that is obtained by replacing the population expectation by the average over subsamples and $\theta$ by $t_n$. If $n$ is large but $m/n$ is small, then random fluctuations in $t_n$ are small relative to those in $t_m$. Accordingly, the sampling distributions of $\rho_m(t_m - t_n)$ and $\rho_m(t_m - \theta)$ are close. Similarly, if $N_{nm}$ is large, the average over subsamples is a good approximation to the population average. These ideas are formalized by Politis and Romano (1994), who show that if $\rho_m / \rho_n$ and $m/n \to 0$ as $m, n \to \infty$ and if $T_n$ has a well-behaved asymptotic distribution, then the non-replacement subsampling method consistently estimates this distribution. The non-replacement subsampling method also consistently estimates asymptotic critical values for $T_n$ and asymptotic confidence intervals for $t_n$. Andrews and Guggenberger (2010) show that non-replacement subsampling is not uniformly consistent in a certain sense. Andrews and Guggenberger (2009) describe methods for overcoming this problem.

In practice, $N_{nm}$ is likely to be very large, which makes $G_{nm}$ hard to compute. This problem can be overcome by replacing the average over all $N_{nm}$ subsamples with the average over a random sample of subsamples (Politis and Romano 1994). These can be obtained by sampling the data $\{X_i : i = 1,...,n\}$ randomly without replacement.

The non-replacement subsampling method enables the asymptotic distributions of statistics to be estimated consistently under very weak conditions. However, the bootstrap is typically more accurate than non-replacement subsampling when the former is consistent. Under conditions that are satisfied in most applications of the bootstrap, the error made by the nonparametric, parametric, and residual bootstrap estimators of a distribution are at most $O_p(n^{-1/2})$ and can be much less. In contrast, the error made by the non-replacement subsampling estimator is $O_p(n^{-1/3})$ (Bugni 2010; Politis and Romano 1994). Bertail (1997); Hall and Jing (1996); and Politis, Romano, and Wolf (1999) describe extrapolation methods for improving this rate, though the improved rate is slower than that of the bootstrap under similar conditions. Thus, the bootstrap estimator of $G_n(\tau, F_0)$ is more accurate than the non-replacement subsampling estimator in most applications in econometrics. The subsampling method is useful, however, however, if characteristics of the sampled population or the statistic of interest cause the bootstrap to be inconsistent or checking the consistency of the bootstrap is difficult.

Another method of subsampling called the $m$ out of $n$ bootstrap consists of drawing $m < n$ observations randomly with replacement from the estimation sample. Except for the size of the bootstrap sample, the $m$ out of $n$ bootstrap is identical to the standard bootstrap. The $m$ out of $n$ bootstrap has properties similar to those of non-replacement subsampling. Swanepoel (1986) gives conditions under which the $m$ out of $n$ bootstrap consistently estimates the distribution of the distribution of the maximum of a sample. Andrews (2000) gives conditions under which it consistently estimates the distribution of a parameter on the boundary of the parameter set. Bickel, Götze, and van Zwet (1997) provide a detailed discussion of the consistency and rates of convergence of the $m$ out of $n$ bootstrap and of an extrapolation method to increase the rate of convergence. Chung and Lee (2001) describe a method for improving the accuracy of the $m$ out of $n$ bootstrap to that of the conventional bootstrap in certain situations.

## 3. ASYMPTOTIC REFINEMENTS

The term "asymptotic refinements" refers to the ability of the bootstrap to provide approximations to the distributions of statistics that are more accurate than the approximations of conventional asymptotic distribution theory. This section explains how the bootstrap provides asymptotic refinements for a large class of statistics that are important in applied research. We continue to assume that the data are an independent random sample from a distribution. Throughout this section and in Sections 4-5, the term "bootstrap" refers to the nonparametric bootstrap of Section 2.1 and the parametric and residual bootstraps of Section 2.2 unless otherwise stated.

The class of statistics treated in this section is called the smooth function model. To define this model, let $\{X_i : i = 1, ..., n\}$ be an independent random sample from the distribution of the random variable or vector $X$. As in Section 2, it is not necessary to distinguish between dependent and explanatory variables in this section. Accordingly, the components of $X$ and the $X_i$'s include any dependent variables. Let $Z(\cdot)$ be a possibly vector valued function on the support of $X$. Define $\theta = EZ(X)$, $Z_i = Z(X_i)$, and $\bar{Z} = n^{-1}\sum_{i=1}^{n} Z_i$. Let $H(z)$ be a "smooth" scalar-valued function whose argument has the same dimension as $Z$. "Smooth" means that $H(z)$ has bounded partial derivatives of sufficiently high order with respect to any combination of the components of $z$. The required order of the derivatives depends on the bootstrap application (e.g., bias reduction, a symmetrical confidence interval) and is not specified here. Let $s_n^2$ be a consistent estimator of $Var\{n^{1/2}[H(\bar{Z}) - H(\theta)]\}$. The statistics in the smooth function model include

1. $H(\bar{Z})$, which estimates $H(\theta)$.

2. $n^{1/2}[H(\bar{Z}) - H(\theta)]$ and $n^{1/2}[H(\bar{Z}) - H(\theta)]/s_n$, which can be used to form confidence intervals for and test hypotheses about $H(\theta)$.

These statistics are all smooth functions $H$ of the sample moments $\bar{Z}$. Many estimators and test statistics used in applied econometrics either are smooth functions of sample moments or can be approximated by such functions with an approximation error that decreases very rapidly as $n$ increases and, is negligible. The ordinary least squares estimator of the slope coefficients in a linear regression model and the $t$ statistic for testing a hypothesis about a coefficient are exact functions of sample moments. Maximum-likelihood and generalized-method-of-moments estimators of the parameters of nonlinear models can be approximated with asymptotically negligible error by smooth functions of sample moments if the log-likelihood function or moment conditions have sufficiently many derivatives with respect to the unknown parameters.

Some test statistics do not satisfy the assumptions of the smooth function model. Quantile estimators, such as the least-absolute-deviations (LAD) estimator, do not satisfy the assumptions of the smooth function model because their objective functions are not sufficiently smooth. Nonparametric density and mean-regression estimators and semiparametric estimators that require kernel or other forms of smoothing also do not fit within the smooth function model. Bootstrap methods for such estimators are discussed in Sections 4 and 5.

*3.1 Bias Reduction*

This section explains how the bootstrap can be used to reduce the finite-sample bias of an estimator. To minimize the complexity of the discussion, it is assumed that $\theta$ and $Z$ in the smooth function model are scalars. However, the method outlined in this section applies to any estimator that satisfies the assumptions of the smooth function model.

To begin, define $\mu = H(\theta)$. Observe that $m_n = H(\bar{Z})$ is a consistent estimator of $\mu$ but $m_n$ is biased if $H$ is a nonlinear function. That is, $m_n \to^{a.s.} \mu$ but $E(m_n) \neq \mu$. A Taylor series expansion of $H(\bar{Z})$ about $\bar{Z} = \theta$ yields

$$H(\bar{Z}) = H(\theta) + H'(\theta)(\bar{Z} - \theta) + (1/2)H''(\theta)(\bar{Z} - \theta)^2 + r_n,$$

where $r_n$ is a remainder term that satisfies $E(r_n) = O(n^{-2})$. Therefore, the bias of $m_n$ is

$$E(m_n - \mu) = (1/2)H''(\theta)E(\bar{Z} - \theta)^2 + O(n^{-2}).$$

The leading term of the bias is

$$B_n \equiv (1/2)H''(\theta)E(\bar{Z} - \theta)^2 = O(n^{-1}).$$

Therefore, the bias of $m_n$ through $O(n^{-1})$ is $B_n$.

Now consider the bootstrap. Let $\{X_i^* : i = 1,...,n\}$ be a bootstrap sample that is obtained by sampling the $X_i$'s randomly with replacement. Define $\bar{Z}^* = n^{-1}\sum_{i=1}^{n} Z(X_i^*)$. The bootstrap analog of $\mu$ is $m_n$, and the bootstrap estimator of $m_n$ is $m_n^* = H(\bar{Z}^*)$. A Taylor series expansion of $H(\bar{Z}^*)$ about $\bar{Z}^* = \bar{Z}$ gives the bootstrap bias through $O(n^{-1})$ almost surely as

$$B_n^* = (1/2)H''(\bar{Z})E^*(m_n^* - m_n)^2,$$

where $E^*$ denotes the expectation under bootstrap sampling, (i.e., the expectation relative to the empirical distribution of the estimation data and conditional on the estimation data). Because the distribution that the bootstrap samples is known, $B_n^*$ can be computed with arbitrary accuracy by Monte Carlo simulation. Thus, $B_n^*$ is a feasible estimator of the bias of $m_n$.

The differences between $B_n$ and $B_n^*$ are that $\bar{Z}$ replaces $\theta$ and $E^*$ replaces $E$ in in $B_n^*$. Moreover, $E(B_n^*) = B_n + O(n^{-2})$. Therefore, through $O(n^{-1})$ the bootstrap bias estimate $B_n^*$ provides the same bias reduction that would be obtained if the infeasible population value $B_n$ could be used. This is the source of the bootstrap's ability to reduce the bias of $m_n$. The resulting bias-corrected estimator of $\mu$ is $m_{n,Corr} = m_n - B_n^*$. It satisfies $E(m_{n,Corr} - \mu) = O(n^{-2})$. Thus, the bias of $m_{n,Corr}$ is $O(n^{-2})$, whereas the bias of $m_n$ is $O(n^{-1})$.

*3.2 Asymptotic Refinements to the Distributions of Test Statistics*

This section explains the bootstrap's ability to provide approximations to the distributions of test statistics that are more accurate than the approximations of conventional asymptotic distribution theory such as the asymptotic normal approximation. We assume that the estimators and test statistics of interest belong to the smooth function model. Specifically, we work with statistics of the form

$$\Delta_n = n^{1/2}[H(\bar{Z}) - H(\theta)] \text{ and}$$

$$t_n = n^{1/2}[H(\bar{Z}) - H(\theta)]/s_n,$$

where $\bar{Z}$ and $\theta$, respectively, are the sample average and population mean of the random variable or vector $Z$; $s_n^2$ is a consistent estimator of the variance of the asymptotic distribution of $\Delta_n$; and $H$ has sufficiently many derivatives with respect to the components of its argument. As was explained in the introduction to Section 3, many estimators and test statistics used in applied econometrics are either smooth functions of sample moments or can be approximated by such functions with a negligible error. We add a further assumption, which is called the Cramér condition.

**Cramér Condition**: Let $\tau$ be a vector of constants with the same dimension as $Z$. Let $i = \sqrt{-1}$. $Z$ satisfies the Cramér condition if

$$\limsup_{\|\tau\| \to \infty} |E \exp(i\tau' Z)| < 1.$$

The Cramér condition is satisfied if $Z$ is continuously distributed (that is, has a conventional probability density) but not if $Z$ is discrete. $Z = Z(X)$ is a function of the observed random variable $X$. The Cramér condition is satisfied even if some components of $X$ are discretely distributed if $Z(X)$ is continuously distributed.

Under the assumptions of the smooth function model and the Cramér condition, $\Delta_n$, $t_n$, and their bootstrap analogs have higher-order asymptotic expansions called Edgeworth expansions. Let $G_n(\cdot, F_0)$ and $G_\infty(\cdot, F_0)$, respectively, denote the distribution and asymptotic distribution functions of either $\Delta_n$ or $t_n$ when $X$ is sampled from a population whose cumulative distribution function is $F_0$. The Edgeworth expansions of $\Delta_n$ and $t_n$ are

(3.1) $\quad G_n(\tau, F_0) = G_\infty(\tau, F_0) + n^{-1/2} g_1(\tau, \kappa_1) + n^{-1} g_2(\tau, \kappa_2) + n^{-3/2} g_3(\tau, \kappa_3) + O(n^{-2})$

uniformly over $\tau$, where the $\kappa$'s are vectors of cumulants through order 4 of the distribution of $Z$; $G_\infty$ is the asymptotic distribution function of $\Delta_n$ or $t_n$; $g_1$ and $g_3$ are even functions of their first arguments; $g_2$ is an odd function of its first argument; and $g_1$, $g_2$, and $g_3$ are differentiable functions of their second arguments. The Edgeworth expansions for the bootstrap analogs of $\Delta_n$ and $t_n$ are

(3.2) $\quad G_n(\tau, F_n) = G_\infty(\tau, F_n) + n^{-1/2} g_1(\tau, \kappa_{n1}) + n^{-1} g_2(\tau, \kappa_{n2}) + n^{-3/2} g_3(\tau, \kappa_{n3}) + O(n^{-2})$

almost surely uniformly over $\tau$, where $F_n$ is the empirical distribution of $X$ and the $\kappa_n$'s are vectors of cumulants of the empirical distribution of $Z$.

To evaluate the accuracy of the bootstrap distribution function $G_n(\tau, F_n)$ as an approximation to the population distribution function $G_n(\tau, F_0)$, subtract (3.2) from (3.1) to obtain

(3.3) $\quad G_n(\tau, F_0) - G_n(\tau, F_n) = G_\infty(\tau, F_0) - G_\infty(\tau, F_n) + n^{-1/2} [g_1(\tau, \kappa_1) - g_1(\tau, \kappa_{n1})]$

$$+ n^{-1} [g_2(\tau, \kappa_2) - g_2(\tau, \kappa_{n2})] + O(n^{-3/2})$$

almost surely uniformly over $\tau$. The leading term on the right-hand side of (3.3) is $G_\infty(\tau, F_0) - G_\infty(\tau, F_n)$, whose size is $O(n^{-1/2})$ almost surely because $F_n - F_0 = O(n^{-1/2})$ almost surely uniformly over the support of $F_0$. Thus, the bootstrap makes an error of size $O(n^{-1/2})$ almost surely, which is the same as the size of the error made by conventional asymptotic approximations. In terms of the rate of convergence of the approximation error to zero, the bootstrap has the same accuracy as conventional approximations.

Now focus on $t_n$. This statistic is asymptotically pivotal, meaning that $G_\infty(\tau, F_0)$ does not depend on $F_0$. Most test statistics are asymptotically pivotal, but most estimators are not. Usually, $G_\infty$ for an asymptotically pivotal statistic is the standard normal distribution function. Asymptotic chi-square statistics are quadratic forms of asymptotically normal statistics and have properties that are straightforward modifications of the properties of asymptotic normal statistics.

When a statistic is asymptotically pivotal, the first term on the right-hand size of (3.3) is zero. The second term is $O(n^{-1})$ almost surely because cumulants can be written as smooth functions of moments and, therefore, $\kappa_1 - \kappa_{n1} = O(n^{-1/2})$. Thus, the error of the bootstrap approximation to the distribution function of an asymptotically pivotal statistic converges to zero more rapidly than the conventional asymptotic approximation. In this sense, the bootstrap is more accurate than conventional asymptotic approximations.

The bootstrap approximation to the symmetrical distribution function $P(|t_n| \leq \tau)$ is even more accurate. Because $G_\infty$ does not depend on $F_0$ or $F_n$, $g_1$ and $g_3$ are even functions, and $g_2$ is an odd function,

(3.4) $\quad P(|t_n| \leq \tau) = G_n(\tau, F_0) - G_n(-\tau, F_0) = \dfrac{2}{n} g_2(\tau, \kappa_2) + O(n^{-2})$

uniformly over $\tau$. The bootstrap analog is

(3.5) $\quad P^*(|t_n^*| \leq \tau) = G_n(\tau, F_n) - G_n(-\tau, F_n) = \dfrac{2}{n} g_2(\tau, \kappa_{n2}) + O(n^{-2})$

almost surely, where $t_n^*$ is the bootstrap analog of $t_n$. Because $\kappa_{n2} - \kappa_2 = O(n^{-1/2})$ almost surely, it follows from (3.4) and (3.5) that

(3.6) $\quad P^*(|t_n^*| \leq \tau) - P(|t_n| \leq \tau) = O(n^{-3/2})$

almost surely. In contrast, the error made by conventional asymptotic approximations such as the normal approximation is $O(n^{-1})$.

In summary, the error in the bootstrap approximation to a one-sided distribution function of an asymptotically pivotal statistic is almost surely $O(n^{-1})$. The error in the bootstrap approximation to a symmetrical distribution function is almost surely $O(n^{-3/2})$. In contrast, the errors made by conventional asymptotic approximations to one-sided and symmetrical distribution functions are $O(n^{-1/2})$ and $O(n^{-1})$, respectively.

Now suppose that the asymptotically pivotal statistic $t_n$ is a statistic for testing a null hypothesis $H_0$ about the sampled population. Let $z_{1-\alpha}$ denote the $1-\alpha$ quantile of the distribution of $|t_n|$. Thus, $P(|t_n| \leq z_{1-\alpha}) = 1-\alpha$. A symmetrical test based on $t_n$ rejects $H_0$ at the $\alpha$ level if $|t_n| > z_{1-\alpha}$. However, $z_{1-\alpha}$ is unknown in most settings. Let $z_{1-\alpha}^*$ be the $1-\alpha$ quantile of the bootstrap distribution of $|t_n^*|$. Then $P^*(|t_n^*| \leq z_{1-\alpha}^*) = 1-\alpha$. The quantity $z_{1-\alpha}^*$ can be estimated with any desired accuracy through Monte Carlo simulation. Therefore, $z_{1-\alpha}^*$ can be treated as known. A simulation algorithm is given at the end of this section. A feasible test rejects $H_0$ if $|t_n| > z_{1-\alpha}^*$. If $H_0$ is correct, the probability that it is rejected is $P(|t_n| > z_{1-\alpha}^*)$. The difference between this probability and the nominal rejection probability of $1-\alpha$ (the error in the rejection probability or ERP) is

$$P(|t_n| > z_{1-\alpha}^*) - P^*(|t_n^*| > z_{1-\alpha}^*) = P(|t_n| \leq z_{1-\alpha}^*) - P^*(|t_n^*| \leq z_{1-\alpha}^*).$$

This difference is not given by (3.6) because $z^*_{1-\alpha}$ is a function of the original data and, therefore, a random variable relative to the population distribution of the data. It is not a random variable relative to the bootstrap distribution. A lengthy argument based on an Edgeworth expansion of the distribution of $|t_n| - z^*_{1-\alpha}$ (see, for example, Hall 1992) shows that if $H_0$ is correct, then

$$P(|t_n| > z^*_{1-\alpha}) - \alpha = O(n^{-2}).$$

Thus, the ERP based on the bootstrap critical value $z^*_{1-\alpha}$ is $O(n^{-2})$. In contrast, the ERP based on the asymptotic critical value (e.g., the $1-\alpha/2$ quantile of the $N(0,1)$ distribution) is $O(n^{-1})$. The ERP based on the bootstrap critical value converges to zero more rapidly than the ERP based on the asymptotic critical value. In samples of practical size, a test based on the bootstrap critical value usually has a smaller ERP than a test based on the asymptotic critical value. The reduction in the ERP can be dramatic. See, for example, Horowitz (1994, 1998a).

A one-sided upper tailed test rejects $H_0$ if $t_n$ is too much greater than zero. A one-sided lower-tailed test rejects $H_0$ if $t_n$ is too much less than zero. The bootstrap versions of the two tests have similar properties, so we consider only the upper-tailed test. Define $\tilde{z}^*_{1-\alpha}$ as the $1-\alpha$ quantile of the bootstrap distribution of $t^*_n$. That is, $P^*(t^*_n \leq \tilde{z}^*_{1-\alpha}) = 1-\alpha$. A one-sided upper-tailed test based on $t_n$ and the bootstrap critical value rejects $H_0$ if $t_n > \tilde{z}^*_{1-\alpha}$. Then if $H_0$ is correct

$$P(t_n > \tilde{z}^*_{1-\alpha}) - \alpha = O(n^{-1}).$$

The ERP in a one-tailed test with a bootstrap critical value is $O(n^{-1})$. In contrast, the error with a conventional asymptotic critical value is $O(n^{-1/2})$ (Hall 1992).

There are, however, circumstances in which the ERP with a bootstrap critical value is $O(n^{-3/2})$. Hall (1992) shows that this is true for a one-sided $t$ test of a hypothesis about a slope (but not intercept) coefficient in a homoskedastic, linear, mean-regression model. Davidson and MacKinnon (1999) show that it is true whenever $t_n$ is asymptotically independent of $g_2(\tilde{z}_{1-\alpha/2}, \kappa_{n2})$, where $\tilde{z}_{1-\alpha/2}$ is the asymptotic $1-\alpha$ quantile of $t_n$. Many familiar test statistics satisfy this condition. Pretorius and Swanepoel (2018) describe a method that combines bootstrap estimates with results obtained from analytic Cornish-Fisher expansions, which are inversions of Edgeworth expansions. The method of Pretorius and Swanepoel (2018) achieves an ERP of $O(n^{-3/2})$ and, in some cases, $O(n^{-2})$. However, its need for analytic Cornish-Fisher expansions makes it intractable for most statistics of interest in econometrics.

The foregoing results also apply to symmetrical and one-sided confidence intervals for a parameter. In the smooth function model, define $\mu = H(\theta)$. Then

$$t_n = n^{1/2}[H(\bar{Z}) - \mu]/s_n.$$

A symmetrical $1-\alpha$ confidence interval for $\mu$ based on the bootstrap critical value $z^*_{1-\alpha}$ is

$$H(\bar{Z}) - n^{-1/2} s_n z^*_{1-\alpha} \leq \mu \leq H(\bar{Z}) + n^{-1/2} s_n z^*_{1-\alpha}.$$

A one-sided upper confidence interval is

$$\mu \leq H(\bar{Z}) + n^{-1/2} s_n \tilde{z}^*_{1-\alpha}.$$

A one-sided lower confidence interval can be defined similarly. The difference between the nominal coverage and true coverage probabilities (errors in the coverage probability or ECPs) of the symmetrical confidence interval is $O(n^{-2})$ and $O(n^{-1})$ for the one-sided interval with the qualifications noted in the previous paragraph. The ECPs based on conventional asymptotic critical values are $O(n^{-1})$ and $O(n^{-1/2})$, respectively, for symmetrical and one-sided confidence intervals.

Tests based on statistics that are asymptotically chi-square distributed behave like symmetrical, two-tailed tests. Therefore, their ERP's under a correct $H_0$ are $O(n^{-1})$ with conventional asymptotic critical values and $O(n^{-2})$ with bootstrap critical values.

The results presented in this section show that the bootstrap reduces the ERPs of hypothesis tests and the ECPs of confidence intervals based on smooth, asymptotically pivotal statistics. These include many asymptotically normal and asymptotically chi-square test statistics that are used for testing hypotheses about the parameters of econometric models. Models that satisfy the required smoothness conditions include linear and nonlinear mean-regression models, error-components mean-regression models for panel data, logit and probit models that have at least one continuously distributed explanatory variable, and tobit models. The smoothness conditions are also satisfied by parametric sample-selection models in which the selection equation is a logit or probit model with at least one continuously distributed explanatory variable. Asymptotically pivotal statistics based on median-regression models do not satisfy the smoothness conditions. Bootstrap methods for such statistics are discussed in Section 4. Statistics based on nonparametric estimators also have different properties. These are treated in Section 5.

The ability of the bootstrap to provide asymptotic refinements for smooth, asymptotically pivotal statistics provides a powerful argument for using them in applications of the bootstrap whenever possible. The bootstrap may be applied to statistics that are not asymptotically pivotal, but it does not provide higher-order approximations to their distributions. Estimators of the structural parameters of econometric models (e.g., slope and intercept parameters, including regression coefficients; standard errors, covariance matrix elements, and autoregressive coefficients) usually are not asymptotically pivotal. The asymptotic distributions of centered structural parameter estimators are often normal with means of zero but have variances that depend on the unknown population distribution of the data. The errors of bootstrap estimates of the distributions of statistics that are not asymptotically pivotal converge to zero at the same rate as the errors made by first-order asymptotic approximations.

Bootstrap hypothesis tests and confidence intervals based on the asymptotically pivotal statistic $t_n$ have smaller ERPs and ECPs than tests and confidence intervals based on conventional asymptotic critical values because the bootstrap estimate of the distribution of $t_n$ is a form of low-order Edgeworth expansion that accounts for skewness and kurtosis of the distribution. Normal and chi-square distributional approximations do not account for skewness or kurtosis. The bootstrap based on an asymptotically pivotal statistic should be used to obtain hypothesis tests or confidence intervals, not standard errors. Standard errors are not asymptotically pivotal, and the bootstrap does not provide asymptotic refinements for them. More importantly, standard errors are related in a simple way to quantiles of the normal distribution and, therefore, to confidence intervals based on the asymptotic normal distribution of $t_n$. However, the bootstrap estimate of the distribution of $t_n$ is non-normal, and there is no simple relation between standard errors and quantiles of non-normal distributions. The use of the bootstrap to estimate standard errors when an asymptotically pivotal statistic is available fails to take advantage the bootstrap's ability to provide asymptotic refinements and results that are numerically more accurate than those provided by asymptotic normal or chi-square approximations.

The following is an algorithm for Monte Carlo computation of symmetrical bootstrap critical values and confidence intervals. The algorithm for one-sided critical values and confidence intervals is similar and not given.

**Monte Carlo Procedure for Computing the Bootstrap Critical Value for Testing a Hypothesis about or Forming a Symmetrical Confidence Interval for a Parameter $\mu$**

Step 1: Use the estimation data to compute a consistent, asymptotically normal estimate of $\hat{\mu}_n$ of $\mu$ and its standard error $s_n$.

Step 2: Generate a bootstrap sample of size $n$ by sampling the empirical distribution of the data randomly with replacement, sampling a parametric estimate of the distribution of the data, or using the residual bootstrap. Estimate $\mu$ and its standard error from the bootstrap sample. Call the results $\mu_n^*$ and $s_n^*$. The bootstrap version of of the asymptotically pivotal statistic $t_n$ is

$$t_n^* = n^{1/2}(\mu_n^* - \hat{\mu}_n)/s_n^*.$$

Step 3: Use the results of many repetitions of Step 2 to compute the empirical distribution of $|t_n^*|$. Set $z_{1-\alpha}^*$ equal to the $1-\alpha$ quantile of this distribution.

Step 4: Reject the hypothesis $H_0: \mu = \mu_0$ at the $1-\alpha$ level if $|n^{1/2}(\hat{\mu}_n - \mu_0)/s_n| > z_{1-\alpha}^*$.

Step 5: A symmetrical $1-\alpha$ confidence interval for $\mu$ is

$$\hat{\mu}_n - n^{-1/2}s_n z_{1-\alpha}^* \leq \mu \leq \hat{\mu}_n + n^{-1/2}s_n z_{1-\alpha}^*.$$

---

## 4. NON-SMOOTH STATISTICS

Some important estimators and test statistics do not satisfy the assumptions of the smooth function model. Quantile estimators and Manski's (1975, 1985) maximum score estimator for a binary response model are two examples. The objective function for quantile estimation has cusps, so its first derivative is discontinuous. The objective function for maximum score estimation is a step function, so its first derivative is either zero or infinity. This section outlines the properties of bootstrap methods for quantile and maximum score estimators.

*4.1 Quantile Estimators*

We restrict attention here to the least absolute deviations (LAD) estimator of a linear median regression model. Similar ideas and methods apply to nonlinear quantile models and estimators for other quantiles. The model is

$$Y = \beta'X + U; \quad P(U \leq 0 | X) = 0.5,$$

where $\beta$ is a vector of constant parameters, $X$ is a vector of explanatory variables, and $U$ is an unobserved random variable. Let $\{Y_i, X_i : i = 1,...,n\}$ be an independent random sample from the distribution of $(Y, X)$. The LAD estimator of $\beta$ is

$$\hat{b} = \arg\min_b \sum_{i=1}^{n} |Y_i - b'X_i|$$

(4.1)
$$= \arg\min_b \sum_{i=1}^{n} (Y_i - b'X_i)[2I(Y_i - b'X_i \geq 0) - 1].$$

Bassett and Koenker (1978) and Koenker and Bassett (1978) give conditions under which $n^{1/2}(\hat{b} - \beta)$ is asymptotically normal with mean zero. The bootstrap estimates the distribution of $n^{1/2}(\hat{b} - \beta)$ consistently (De Angelis, Hall, and Young 1993; Hahn 1995), but the non-smoothness of the LAD objective function causes the difference between bootstrap approximation to the distribution function of $n^{1/2}(\hat{b} - \beta)$ and the true distribution function to converge to zero slowly. Depending on how the bootstrap is implemented and certain other conditions, the rate is either $O(n^{-1/4})$ or $O(n^{-2/5})$ (De Angelis, Hall, and Young 1993). In addition, the Edgeworth expansion of the distribution of $n^{1/2}(\hat{b} - \beta)$ is very complicated. It is not known whether bootstrap sampling applied to (4.1) provides asymptotic refinements for hypothesis tests and confidence intervals.

Horowitz (1998b) proposed smoothing the cusp in the LAD objective function to make it differentiable. To state the smoothed objective function, let $H(\cdot)$ be a function with sufficiently many continuous derivatives such that $H(v) = 1$ if $v \geq 1$ and $H(v) = 0$ if $v \leq -1$. For example, $H(v)$ might be the cumulative distribution function of a random variable whose support is $[-1, 1]$. Let $\{h_n\}$ be a decreasing sequence of positive constants that converges to zero as $n \to \infty$. The smoothed LAD (SLAD) estimator is

(4.2) $\quad \tilde{b} = \arg\min_b \sum_{i=1}^{n} (Y_i - b'X_i) \left[ 2H\left( \frac{Y_i - b'X_i}{h_n} \right) - 1 \right].$

The objective function (4.2) is identical to the LAD objective function (4.1) if $|Y_i - b'X_i| \geq h_n$, but it smooths the discontinuities in the indicator function at $Y_i - b'X_i = 0$. The interval over which smoothing occurs decreases as $h_n \to 0$. This is necessary to make $\tilde{b}$ a consistent estimator of $\beta$.

Because the objective function in (4.2) is differentiable, standard Taylor series methods can be used to obtain the asymptotic distribution of $n^{1/2}(\tilde{b} - \beta)$. Horowitz (1998b) gives conditions under which $n^{1/2}(\tilde{b} - \beta) \to^d N(0, V)$ where $V$ is a covariance matrix. Let $\hat{V}$ be a consistent estimator of $V$, $\hat{V}_{jj}$ denote the $(j, j)$ component of $\hat{V}$, and $\tilde{b}_j$ and $\beta_j$ denote the $j$'th components of $\tilde{b}$ and $\beta$, respectively. Then a $t$ statistic for testing the hypothesis $H_0: \beta_j = \beta_{0j}$ is

$$t_n = \frac{n^{1/2}(\tilde{b}_j - \beta_{0j})}{\hat{V}_{jj}^{1/2}}.$$

A bootstrap version of $t_n$ can be obtained by replacing the original sample $\{Y_i, X_i : i = 1,...,n\}$ with the bootstrap sample $\{Y_i^*, X_i^* : i = 1,...,n\}$ in (4.2). Let $\tilde{b}^*$ denote the estimate of $\beta$ obtained from the bootstrap sample and $V_{jj}^*$ denote the bootstrap analog of $\hat{V}_{jj}$. The bootstrap analog of $t_n$ is

$$t_n^* = \frac{n^{1/2}(\tilde{b}_j^* - \tilde{b}_j)}{(V_{jj}^*)^{1/2}}.$$

Horowitz (1998b) shows that $t_n$ and $t_n^*$ have Edgeworth expansions that are identical almost surely through $O[(nh_n)^{-1}]$. Therefore, arguments similar to those in Section 3.2 show that the bootstrap provides asymptotic refinements for hypothesis tests and confidence intervals based on the SLAD estimator. For example, consider a symmetrical $t$ test of $H_0$. Let $z_{1-\alpha}^*$ be the $1-\alpha$ quantile of the bootstrap distribution of $|t_n^*|$. Then

$$P(|t_n| > z_{1-\alpha}^*) = \alpha + o[(nh_n)^{-1}]$$

if $H_0$ is true. In contrast, the asymptotic normal approximation makes an error of $O[(nh_n)^{-1}]$. Use of the bootstrap critical value reduces the approximation error because the bootstrap captures a higher order term in the Edgeworth expansion of $|t_n|$, whereas the asymptotic normal approximation drops this term. The bootstrap also provides asymptotic refinements for one-sided hypothesis tests and confidence intervals and for asymptotic chi-square tests of hypotheses about several components of $\beta$. In addition, the bootstrap provides asymptotic refinements for a smoothed version of Powell's (1984, 1986) censored LAD estimator.

*4.2 The Maximum Score Estimator*

Manski's (1975, 1985) maximum score estimator applies to the binary response model $Y = I(\beta'X + U \geq 0)$, where $Y$ is an observed random variable, $X$ is an observed random vector, $U$ is an unobserved random variable satisfying $P(U \leq 0 \mid X) = 0.5$, and $\beta$ is an unknown vector of constant parameters to be estimated. The vector $\beta$ is identified only up to scale, so a scale normalization is needed. Here, scale normalization will consist of setting $|\beta_1| = 1$, where $\beta_1$ is the first component of $\beta$. Assume that the first component of $X$ is continuously distributed. Define the set $B = \{b : |b_1| = 1\}$. Let $\{Y_i, X_i : i = 1, ..., n\}$ be an independent random sample of $(Y, X)$. The maximum score estimator of $\beta$ is

(4.3) $\hat{b}_{MS} = \arg\max_{b \in B} \sum_{i=1}^{n} (2Y_i - 1) I(b'X_i \geq 0)$.

Manski (1975, 1985) gives conditions under which $\hat{b}_{MS} \to \beta$ almost surely. Cavanagh (1987) and Kim and Pollard (1990) show that $\hat{b}_{MS}$ converges to $\beta$ at the rate $n^{-1/3}$ and that $n^{1/3}(\hat{b}_{MS} - \beta)$ has a complicated, non-normal asymptotic distribution. Abrevaya and Huang (2005) show that the nonparametric bootstrap does not provide a consistent estimate of this distribution.

The maximum score estimator converges slowly and has a complicated limiting distribution because it is obtained by maximizing a step function. Several authors have proposed bootstrap sampling procedures and/or modifications of the maximum score objective function that make the bootstrap consistent (Cattaneo, Jansson, and Nagasawa 2018; Hong and Li 2015; Horowitz 1992, 2002; Patra, Seijo, and Sen 2018). The method of Horowitz (1992, 2002), which provides asymptotic refinements, is described in this section. Horowitz (1992) proposed replacing the indicator function on the right-hand side of (4.3) with a differentiable function. Let $H$ and $h_n$ be defined as in (4.2). The resulting smoothed maximum score (SMS) estimator is

$$\hat{b}_{SMS} = \arg\max_{b \in B} \sum_{i=1}^{n} (2Y_i - 1) H\left(\frac{b'X_i}{h_n}\right).$$

Horowitz (1992) gives conditions under which $\hat{b}_{SMS}$ converges to $\beta$ at a rate that is at least as fast as $n^{-2/5}$ and $(nh_n)^{1/2}(\hat{b}_{SMS} - \beta)$ is asymptotically normally distributed. Asymptotic normality makes it possible to form asymptotically pivotal $t$ statistics for testing hypotheses about $\beta$ and obtain asymptotic refinements with the bootstrap (Horowitz 2002). Let $t_n$ be a $t$ statistic, and let $t_n^*$ be the bootstrap analog that is obtained by smoothed maximum score estimation based on the bootstrap sample $\{Y_i^*, X_i^* : i = 1, ..., n\}$. Now consider a symmetrical $t$ test of the hypothesis $H_0 : \beta_j = \beta_{0j}$. Let $z_{1-\alpha}^*$ be the $1-\alpha$ quantile of the bootstrap distribution of $|t_n^*|$. Then

$$P(|t_n| > z_{1-\alpha}^*) = \alpha + o[(nh_n)^{-1}]$$

if $H_0$ is true. In contrast, the asymptotic normal approximation makes an error of $O[(nh_n)^{-1}]$. As in smoothed LAD estimation, the bootstrap captures a higher order term in the Edgeworth expansion of $|t_n|$, whereas the asymptotic normal approximation drops this term. Also as in smoothed LAD estimation, the bootstrap also provides asymptotic refinements for one-sided hypothesis tests and confidence intervals and for asymptotic chi-square tests of hypotheses about several components of $\beta$.

# 5. NONPARAMETRIC ESTIMATION

This section is concerned with inference about the unknown function $g$ in the nonparametric mean regression model

(5.1) $\quad Y = g(X) + \varepsilon; \quad E(\varepsilon \mid X) = 0$

and the nonparametric quantile regression model

(5.2) $\quad Y = g(X) + \varepsilon; \quad P(\varepsilon \leq 0 \mid X) = \tau; \quad 0 < \tau < 1$.

Nonparametric estimators of $g$ and their properties are described by Fan and Gijbels 1996; Fan, Hu, and Truong (1994); Härdle (1990); and Yu and Jones (1997) among many others.

Any method for testing a hypothesis about $g$ or constructing a confidence interval or band for $g$ based on a nonparametric estimate must deal with the problem of asymptotic bias. Let $\hat{g}$ denote a nonparametric estimate of $g$. The expected value of $\hat{g}$ does not equal $g$, the asymptotic normal distribution of the scaled estimate is not centered at $g$, and the true coverage probability of an asymptotic confidence interval for $g$ (or rejection probability of a hypothesis test) that is constructed from the normal distribution in the usual way is less than the nominal probability. This problem is usually overcome by undersmoothing or explicit bias reduction. Undersmoothing consists of making the bias asymptotically negligible by using a bandwidth whose rate of convergence is faster than the asymptotically optimal rate. In explicit bias reduction, an estimate of the asymptotic bias is used to construct an asymptotically unbiased estimate of $g$. Most explicit bias reduction methods involve some form of oversmoothing, that is using a bandwidth whose rate of convergence is slower than the asymptotically optimal rate. Hall (1992) and Horowitz (2001) describe the ability of the bootstrap to provide asymptotic refinements for hypothesis tests and confidence intervals based on nonparametric estimators with undersmoothing or explicit bias correction.

Methods based on undersmoothing or oversmoothing require a bandwidth whose rate of convergence is faster or slower than the asymptotically optimal rate. However, there are no effective empirical ways to choose these bandwidths. Hall and Horowitz (2013) (HH) and Horowitz and Krishnamurthy (2018) (HK) describe bootstrap methods for overcoming this problem in models (5.1) and (5.2), respectively. These methods use bandwidths chosen by standard empirical methods such as cross validation or a plug-in rule. Instead of under- or oversmoothing, the methods use the bootstrap to estimate the bias of $\hat{g}$. The bootstrap estimate of the bias can be obtained using the procedure of Section 3.1. However, in this nonparametric setting, the bootstrap bias estimate has stochastic noise that is comparable in size to the bias itself. HH and HK give conditions under which combining a suitable quantile of the "distribution" of the bootstrap bias estimate with $\hat{g}$ provides pointwise confidence intervals and uniform confidence bands with asymptotic coverage probabilities that equal or exceed the nominal probabilities. The methods of HH and HK use the bootstrap to select critical values that are larger than those obtained from the usual asymptotic normal approximations and, therefore, yield wider confidence intervals that achieve the desired coverage probabilities asymptotically without the need for bias correction through undersmoothing or explicit bias estimation. The same methods can also be used to form hypothesis tests. In contrast to the use of the bootstrap in Sections 3 and 4, the objective here is not to obtain asymptotic refinements. Rather, the bootstrap is used to overcome a problem in constructing first-order asymptotic confidence bands and hypothesis tests.

HH and HK provide the details of the methods, which are lengthy, though not difficult to implement. As an illustration, the method for obtaining a pointwise confidence band for the mean-regression model (5.1) with a scalar covariate $X$ with compact support $\mathcal{R}$ and a homoskedastic $\varepsilon$ consists of the following steps.

<u>Step 1</u>: Form an estimate $\hat{g}$ of $g$ by using a kernel-based method such as a local polynomial estimator. Use a standard empirical method such as cross-validation or plug-in to choose the bandwidth. Form an estimate $\hat{\sigma}^2$ of $\sigma^2 = E(U^2)$ by using a standard method such as that of Rice (1984). Let $\mathcal{S}(x)\hat{\sigma}^2$ denote an estimate of the variance of $\hat{g}(x)$. In local polynomial estimation, $\mathcal{S}(x)$ is a known function of the $X_i$'s in the data. HH provide details.

<u>Step 2</u>: Compute residuals $\tilde{\varepsilon}_i = Y_i - \hat{g}(X_i)$ ($i = 1,...,n$). Set $\bar{\varepsilon} = n^{-1}\sum_{i=1}^{n}\tilde{\varepsilon}_i$, and compute centered residuals $\hat{\varepsilon}_i$ defined by $\hat{\varepsilon}_i = \tilde{\varepsilon}_i - \bar{\varepsilon}$.

<u>Step 3</u>: Form the residual bootstrap sample $\{Y_i^*, X_i : i = 1,...,n\}$, where $Y_i^* = \hat{g}(X_i) + \varepsilon_i^*$ and the $\varepsilon_i^*$'s are obtained by sampling the $\hat{\varepsilon}_i$'s randomly with replacement. The $X_i$'s are the same as in the original data and are not resampled.

<u>Step 4</u>: Use the estimation methods and bandwidth of Step 1 with the bootstrap sample from Step 3 to obtain bootstrap estimates of $g$ and $\sigma^2$. Denote these by $g^*$ and $\sigma^{*2}$. This step uses the bandwidth obtained in Step 1. It does not use the bootstrap sample to obtain a new bandwidth. Form the bootstrap pointwise $1-\alpha$ confidence band

$$\mathcal{B}^*(\alpha) = \{(y,x): g^*(x) - \mathcal{S}(x)\sigma^* z_{1-\alpha/2} \leq \hat{g}(x) \leq g^*(x) + \mathcal{S}(x)\sigma^* z_{1-\alpha/2}\},$$

where $z_{1-\alpha/2}$ is the $1-\alpha/2$ quantile of the $N(0,1)$ distribution.

<u>Step 5</u>: Repeat Step 4 $B$ times, where $B$ is a large positive integer (e.g., $B = 1000$). Let $\mathcal{B}_b^*(\alpha)$ denote the confidence band obtained on the $b$'th out of $B$ repetition. For $x \in \text{supp}(X)$, compute the bootstrap coverage probability

$$\pi^*(x,\alpha) = B^{-1}\sum_{b=1}^{B} I[(x,\hat{g}(x)) \in \mathcal{B}_b^*(\alpha)].$$

<u>Step 6</u>: Let $1-\alpha_0$ be the desired coverage probability of a pointwise confidence band. Define $\beta^*(x,\alpha_0)$ to be the solution in $\alpha$ of $\pi^*(x,\alpha) = 1-\alpha_0$. For $\xi \in (0,.5]$, let $\alpha_\xi^*(\alpha_0)$ denote the $\xi$-level quantile of points in the set $\{\beta^*(x,\alpha_0): x \in \mathcal{R}\}$. In practice, it suffices to find the quantile over a closely space grid of discrete points in $\mathcal{R}$. Asymptotically, the confidence band

$$\mathcal{B}[\alpha_\xi(\alpha_0)] = \{(y,x): \hat{g}(x) - \mathcal{S}(x)\hat{\sigma}z_{1-\alpha_\xi(\alpha_0)/2} \leq g(x) \leq \hat{g}(x) + \mathcal{S}(x)\hat{\sigma}z_{1-\alpha_\xi(\alpha_0)/2}\}$$

covers all but a fraction $1-\xi$ of points in $x \in \mathcal{R}$. The exceptional points are in regions where $g(x)$ has sharp peaks or troughs that cause the bias of $\hat{g}$ to be unusually large. These regions are typically visible in a plot of $\hat{g}$. HH provide a theoretical analysis of this phenomenon and Monte Carlo illustrations of the numerical performance of the method for the mean regression model (5.1). HK illustrate the performance of the method for the quantile regression model (5.2).

## 6. THE MULTIPLIER AND WILD BOOTSTRAPS

The multiplier and wild bootstraps are methods for generating bootstrap samples that do not consist of resampling the original data or residuals as in Sections 2-5. Rather, the multiplier and wild bootstraps combine the data with random variables drawn from a known distribution to form a bootstrap sample. The multiplier and wild bootstraps provide ways to deal with issues such as heteroskedasticity of unknown form in fixed-design regression models or random-design models in which one conditions on the covariates. They also provide methods for obtaining non-asymptotic bounds on the ERPs of certain hypothesis tests.

*6.1 The Wild Bootstrap*

The wild bootstrap enables accurate inference to be carried out in regression models with heteroskedasticity of unknown form. This section focusses on the linear model

(6.1) $Y_i = \beta' X_i + \varepsilon_i; \ i = 1,...,n$

where the $X_i$'s are fixed in repeated samples or random but inference is conditional on the observed $X_i$'s; $E(\varepsilon_i) = 0$ for all $i = 1,...,n$; $E(\varepsilon_i^2) = \sigma^2(X_i)$; and $\sigma^2(\cdot)$ is an unknown function. Asymptotic inference about $\beta$ can be carried out by using the heteroskedasticity-consistent covariance matrix estimator (HCCME) of Eicker (1963, 1967) and White (1980), but this estimator can be severely biased downward in finite samples with the consequence that the true finite-sample probability of rejecting a correct null hypothesis can be significantly larger than the nominal probability (Chesher and Jewitt 1987). If the $X_i$'s are treated as random, the bootstrap can be implemented for model (6.1) by sampling the data $\{Y_i, X_i : i =,...,n\}$ randomly with replacement. However, this method can be inaccurate because it does not impose the moment condition $E(\varepsilon_i | X_i) = 0$.

The wild bootstrap imposes the moment restriction. It was introduced by Liu (1988) following a suggestion of Wu (1986). Mammen (1993) establishes the ability of the wild bootstrap to provide asymptotic refinements for the model (6.1). Cao-Abad (1991), Härdle and Mammen (1993), and Härdle and Marron (1991) use the wild bootstrap in nonparametric regression. To describe the method for a linear model, let $\hat{b}$ be the ordinary least squares (OLS) estimate of the vector of slope coefficients, $\beta$, based on data $\{Y_i, X_i : i = 1,...,n\}$ in the model

(6.2) $Y_i = \beta' X_i + \varepsilon_i; \ E(\varepsilon_i = 0); \ E(\varepsilon_i^2) = \sigma_i^2$.

The variances $\sigma_i^2$ are unknown and not necessarily equal. The wild bootstrap generates bootstrap samples $\{Y_i^*, X_i : i = 1,...,n\}$ from

(6.3) $\quad Y_i^* = \hat{b}'X_i + \varepsilon_i^*.$

As in the residual bootstrap method of Section 2.2, the wild bootstrap uses the $X_i$'s from the original data. The $X_i$'s are not resampled. The $\varepsilon_i^*$'s are generated by either of the following two methods:

1. Let $\hat{\varepsilon}_i = Y_i - \hat{b}'X_i$ ($i = 1,...,n$) be the OLS residuals from model (6.2). For each $i = 1,...,n$, let $F_i$ be the unique 2-point distribution that satisfies $E(\eta_i) = 0$; $E(\eta_i^2 | F_i) = \hat{\varepsilon}_i^2$; and $E(\eta_i^3) = \hat{\varepsilon}_i^3$; where $\eta_i$ is a random variable with the cumulative distribution function $F_i$. Then, $\eta_i = (1 - \sqrt{5})\hat{\varepsilon}_i$ with probability $(1 + \sqrt{5})/(2\sqrt{5})$, and $\eta_i = (1 + \sqrt{5})\hat{\varepsilon}_i / 2$ with probability $1 - (1 + \sqrt{5})/(2\sqrt{5})$. Set $\varepsilon_i^* = \eta_i$ for each $i$. Mammen (1993) provides a detailed discussion of the properties of this method.

2. This method is an example of the multiplier bootstrap, meaning that the $\varepsilon_i^*$'s are multiples of transformations of the residuals $\hat{\varepsilon}_i$ and independent random variables. Specifically, let $U_i$ ($i = 1,...,n$) be random variables that are independent of each other and the OLS residuals such that $E(U_i) = 0$ and $E(U_i^2) = 1$. One possibility is $U_i \sim N(0,1)$. Let $f(\hat{\varepsilon}_i)$ be a transformation of the OLS residuals, possibly $f(\hat{\varepsilon}_i) = \hat{\varepsilon}_i$. Set $\varepsilon_i^* = U_i f(\hat{\varepsilon}_i)$. Davidson and Flachaire (2008) discuss properties of this method.

Regardless of the method used to generate $\varepsilon_i^*$, implementation of the wild bootstrap proceeds as follows:

Step 1: Generate a bootstrap sample $\{Y_i^*, X_i : i = 1,...,n\}$ from (6.3). Estimate $\beta$ by OLS using this sample. Compute the resulting bootstrap $t$ statistic, $t_n^*$ by using the HCCME or a variant such as that of MacKinnon and White (1985). .

Step 2: Obtain the empirical distribution of $t_n^*$ by repeating step 1 many times. Obtain the critical value of $t_n^*$. Use this critical value with the $t$ statistic from the original data to test hypotheses about and form confidence intervals for the components of $\beta$.

Horowitz (2001) and Davidson and Flachaire (2008) provide examples of the numerical performance of the wild bootstrap.

*6.2 Non-Asymptotic Inference in Maximum Likelihood Estimation*

This section describes the use of the multiplier bootstrap to carry out a likelihood ratio test of a finite-dimensional parameter $\theta \in \mathbb{R}^d$ for some finite integer $d \geq 1$. The hypothesis is $H_0 : \theta = \theta_0$ for some known $\theta_0$. Let the data $\{X_i : i = 1,...,n\}$ be an independently but not necessarily identically distributed random sample from some population.. As in Section 2, the components of $X_i$ include any dependent variables as well as explanatory variables. Let $f_i(\cdot, \theta)$ denote the probability density function of the $i$'th observation. Technically, $f_i$ is a Radon-Nikodym density, so the some components of $X_i$ may be discretely distributed or have probability distributions that are continuous in some regions and discrete in others, but this section uses ordinary probability density notation to minimize the complexity of the discussion. The log-likelihood function is

$$\log L(\theta) = \sum_{i=1}^{n} \log f_i(X_i, \theta).$$

The maximum likelihood estimate of $\theta$ is

$$\hat{\theta} = \arg\max_{\theta \in \Theta} \log L(\theta),$$

where $\Theta \subset \mathbb{R}^d$ is the parameter set. Under $H_0: \theta = \theta_0$,

$$\theta_0 = \arg\max_{\theta \in \Theta} E[\log L(\theta)].$$

The likelihood ratio statistic for testing $H_0$ is

$$LR = 2[\log L(\hat{\theta}) - \log L(\theta_0)].$$

To construct the multiplier bootstrap version of $LR$, let $\{U_i : i = 1,...,n\}$ be scalar random variables that are independent of each other and the $X_i$'s such that $E(U_i) = E(U_i^2) = 1$ and $E\exp(U_i) < \infty$ for all $i = 1,...,n$. The multiplier bootstrap version of the log-likelihood function is

$$\log L^*(\theta) = \sum_{i=1}^{n} U_i \log f_i(X_i, \theta).$$

The multiplier bootstrap parameter estimate is

$$\theta^* = \arg\max_{\theta \in \Theta} \log L^*(\theta).$$

Let $E^*$ denote the expectation with respect to the distribution of the $U_i$'s with the $X_i$'s held constant. Then under $H_0$

$$\theta_0 = \arg\max_{\theta \in \Theta} E^*[\log L^*(\theta)],$$

and the multiplier bootstrap version of the likelihood ratio statistic is

$$LR^* = 2[\log L(\theta^*) - \log L(\theta_0)].$$

Let $P_n^*$ denote the probability distribution induced by sampling the $U_i$'s while holding the $X_i$'s constant. Let $z_{1-\alpha}^*$ denote the $1-\alpha$ quantile of the distribution of $LR^*$ under $P_n^*$ probability. That is

$$z^*_{1-\alpha} = \inf_{z \geq 0}\{z: P^*_n(LR^* > z) \leq \alpha\}.$$

The quantity $z^*_{1-\alpha}$ can be estimated with any desired accuracy by repeated sampling of the $U_i$'s and computation of $LR^*$. Spokoiny and Zhilova (2015) give condtions under which

$$(6.4) \quad \left|P\left(LR > z^*_{1-\alpha}\right) - \alpha\right| \leq C\left(\frac{d+v}{n}\right)^{1/8}$$

for any $v$ such that $\alpha \leq 1 - 8e^{-v}$, where $C$ is a constant. The right-hand side of (6.4) is small only if $n$ is very large. Nonetheless, (6.4) is important because it is a finite-sample result showing that a version of the bootstrap can be used to carry out non-asymptotic inference with a familiar and frequently used statistic. In addition, the right-hand side of (6.4) is a worst case bound that accommodates the most extreme distributions of $\{X_i : i = 1,...,n\}$. Spokoiny and Zhilova (2015) provide Monte Carlo evidence indicating that the bound is much tighter than (6.4) suggests in less extreme cases.

*6.3 Inference with the Multiplier Bootstrap in High-Dimensional Settings*

Let $\{X_i : i = 1,...,n\}$ be independent $p \times 1$ random vectors with means of zero and finite covariance matrices. Define the $p \times 1$ random vector

$$X = n^{-1/2}\sum_{i=1}^{n} X_i,$$

and let $X^{(j)}$ denote the $j$'th component of $X$. Chernozhukov, Chetverikov, and Kato (2013) (CCK) consider the problem of estimating the probability distribution of

$$Z = \max_{1 \leq j \leq p} X^{(j)}$$

when $p$ may be greater than $n$. This problem arises in testing multiple hypotheses and certain high-dimensional estimation methods, among other applications.

If $p$ is fixed and certain other conditions are satisfied, a standard central limit theorem shows that $X$ is asymptotically multivariate normal. The asymptotic distribution of $Z$ can be obtained from this result. However, standard central limit theorems do not apply if $p > n$.

CCK define a multiplier bootstrap version of $Z$. Let $\{e_i : i = 1,...,n\}$ denote a sequence of $N(0,1)$ random variables that are independent of each other and the $X_i$'s. Let $X^{*(j)}$ be the $j$'th component of the $p \times 1$ random vector

$$X^* = n^{-1/2} \sum_{i=1}^{n} X_i e_i.$$

Define the following multiplier bootstrap version of $Z$:

$$Z^* = \max_{1 \leq j \leq p} n^{-1/2} X^{*(j)}.$$

Let $z^*_{1-\alpha}$ denote the $1-\alpha$ quantile of the distribution of $Z^*$ under sampling of the $e_i$'s while holding the $X_i$'s fixed at the values in the data. CCK give conditions under which $|P(X \leq z^*_{1-\alpha}) - (1-\alpha)| \to 0$ as $n \to \infty$ even if $p > n$. CCK also show that their version of the multiplier bootstrap provides consistent estimates of the distributions of statistics related to $Z$. Finally, CCK give examples of the application of their results to certain problems in high-dimensional estimation and testing.

## 7. THE LASSO

The LASSO (Least Absolute Shrinkage and Selection Operator) is a method for parameter estimation in settings in which the number of unknown parameters is comparable to or may exceed the sample size. The LASSO was introduced by Tibshirani (1996). It has been the object of much subsequent research and has generated a vast literature. Bühlmann and van de Geer (2011) synthesize the results of this research. The LASSO is applicable to a wide variety of models. Here, however, we treat only the homoskedastic linear model

(7.1) $Y_i = \sum_{j=1}^{p} \beta_j X_{ij} + \varepsilon_i; \quad i = 1,...,n$,

where $\{\beta_j : j = 1,...,p\}$ are constant parameters to be estimated, $\{X_{ij} : i = 1,...,n; j = 1,...,p\}$ are covariates that may be fixed in repeated samples or random, and $\{\varepsilon_i : i = 1,...,n\}$ are unobserved random variables that are independently and identically distributed with means of zero and finite variances. Model (7.1) is assumed to be sparse in the sense that most of the $\beta_j$'s are zero or close to zero in a certain sense but a relatively small number are non-zero. It is possible that $p > n$, in which case the $\beta_j$'s cannot be estimated by OLS.

The LASSO consists of estimating the $\beta_j$'s by solving the penalized least squares problem

$$\hat{b} \equiv (\hat{b}_1,...,\hat{b}_p)' = \arg\min_b \sum_{i=1}^{n}\left(Y_i - \sum_{j=1}^{p} b_j X_{ij}\right)^2 + \lambda \sum_{j=1}^{p} |b_j|,$$

where $\lambda > 0$ is a constant penalization parameter. This estimator has a complicated, non-normal asymptotic distribution even if $p < n$ (Knight and Fu 2000). The bootstrap does not estimate this distribution consistently (Chatterjee and Lahiri 2011).

The adaptive Lasso (Zou 2006) is a variant of the LASSO that overcomes these problems. The adaptive LASSO (ALASSO) estimates $\beta$ in two steps. The first step consists of obtaining a $n^{-1/2}$-consistent initial estimate of $\beta$. For example, $\tilde{b}$ can be the OLS estimate if $p < n$ and the LASSO estimate if $p > n$. Denote the initial estimate by $\tilde{b} = (\tilde{b}_1,...,\tilde{b}_p)$. The second step estimate is

$$\hat{b}_{AL} = \arg\min_{b} \sum_{i=1}^{n}\left(Y_i - \sum_{\substack{j=1 \\ \tilde{b}_j \neq 0}}^{p} b_j X_{ij}\right)^2 + \lambda \sum_{\substack{j=1 \\ \tilde{b}_j \neq 0}}^{p} \frac{|b_j|}{|\tilde{b}_j|},$$

where $\hat{b}_{AL,j} = 0$ if $\tilde{b}_j = 0$. The ALASSO estimate is asymptotically normal. In particular, let number of non-zero components of $\beta$ be fixed. Define $A_+ = \{j: \beta_j \neq 0\}$, $\beta_+ = \{\beta_j : j \in A_+\}$, and $\hat{b}_{AL+} = \{\hat{b}_{AL,j} : j \in A_+\}$. Then $n^{1/2}(\hat{b}_{AL+} - \beta_+) \to^d N(0, V_+)$, where $V_+$ is the covariance matrix of the OLS estimate of $\beta_+$ obtained from a model that contains only explanatory variables $X_{ij}$ for which $j \in A_+$. This property is called oracle efficiency.

To implement the bootstrap with the ALASSO, let $\tilde{\varepsilon}_i = Y_i - \hat{b}'_{AL} X_i$ ($i = 1,...,n$) be the ALASSO residuals, $\bar{\varepsilon} = n^{-1} \sum_{i=1}^{n} \tilde{\varepsilon}_i$, and $\hat{\varepsilon}_i = \tilde{\varepsilon}_i - \bar{\varepsilon}$ be the centered residuals. Define

$$Y_i^* = \sum_{j=1}^{p} \hat{b}_{AL,j} X_{ij} + \varepsilon_i^*; \quad i = 1,...,n,$$

where the $\varepsilon_i^*$'s are drawn randomly with replacement from the $\hat{\varepsilon}_i$'s. The bootstrap sample is $\{Y_i^*, X_i : i = 1,...,n\}$. Let $b_{AL}^*$ denote the AL estimate of $\beta$ based on the bootstrap sample. Chatterjee and Lahiri (2013) give conditions under which the bootstrap provides asymptotic refinements for the ALASSO. As an illustration of their results, let $c$ be a $p \times 1$ vector of constants. Let $t_n$ and $t_n^*$, respectively, be $t$ statistics for testing the hypothesis $c'\beta = 0$ and its bootstrap analog $c\hat{b}_{AL} = 0$. Specifically,

$$t_n = \frac{n^{1/2} c'(\hat{b}_{AL} - \beta)}{s_n}$$

and

$$t_n^* = \frac{n^{1/2} c'(b_{AL}^* - \hat{b}_{AL})}{s_n^*},$$

where $s_n$ is a consistent estimate of the standard error of $n^{1/2} c' \hat{b}_{AL}$ and $s_n^*$ is its bootstrap analog. Let $P^*$ denote probability under bootstrap sampling of the $\varepsilon_i^*$'s conditional on the $X_{ij}$'s. Chatterjee and Lahiri (2013) give conditions under which

$$\sup_\tau | P(t_n \leq \tau) - P^*(t_n^* \leq \tau) | = o_p(n^{-1/2}).$$

This is more accurate than the error made by the asymptotic normal approximation, which is $O(n^{-1/2})$. Chatterjee and Lahiri (2013) also give conditions under which the bootstrap estimates the distribution of a modified $t$ statistic with an accuracy of $O_p(n^{-1})$.

## 8. TIME SERIES DATA

Bootstrap sampling with time series data must capture the dependence of the data-generation process (DGP) in a suitable way. Methods for doing this depend on what assumes about the form of the dependence and can be complex if one assumes little. This section summarizes several methods for implementing the bootstrap with time series data. The details of some methods are lengthy. More thorough presentations are provided in Härdle, Horowitz, and Kreiss (2003); Horowitz (2001); and the references therein and in this section.

Bootstrap sampling can be carried out relatively easily if there is a parametric model, such as an ARMA model, that reduces the DGP to a transformation of independent random variables. For example, suppose that the time series $\{X_t : t = 1,...,T\}$ is generated by the stationary, invertible, finite-order ARMA model

(8.1) $\quad A(L,\alpha) X_t = B(L,\beta) U_t$,

where $A$ and $B$ are known functions, $L$ is the backshift operator, $\alpha$ and $\beta$ are vectors of parameters, and $\{U_t\}$ is a sequence of independently and identically distributed random variables with means of zero. Let $\hat{\alpha}$ and $\hat{\beta}$ be $T^{1/2}$-consistent, asymptotically normal estimators of $\alpha$ and $\beta$, and let $\{\hat{U}_t\}$ be the residuals of the estimated model (8.1) centered so that $T^{-1}\sum_{t=1}^{T}\hat{U}_t = 0$. Then a bootstrap sample $\{X_t^*\}$ can be generated as

$$A(L,\hat{\alpha})X_t^* = B(L,\hat{\beta})U_t^*,$$

where $\{U_t^*\}$ is an independent random sample from the empirical distribution of the centered residuals $\{\hat{U}_t\}$. If the distribution of $U_t$ is assumed to belong to a known parametric family (e.g., the normal distribution), then $\{U_t^*\}$ can be generated by independent sampling from the estimated distribution. Bose (1988) provides a rigorous discussion of the use of the bootstrap with autoregressions. Bose (1990) treats moving average models.

Another possibility is that the data are generated by a stationary, linear process. That is, the DGP has the form

(8.2) $\quad X_i - \mu = \sum_{j=1}^{\infty} \alpha_j (X_{i-j} - \mu) + U_i,$

where $\mu = E(X_i)$ for all $i$, $\{U_i\}$ is a sequence of independently and identically distributed random variables, and $\{X_i\}$ may be a scalar or a vector process. Assume that $\sum_{j=1}^{\infty} \alpha_j^2 < \infty$ and that all of the roots of the power series $1 - \sum_{j=1}^{\infty} \alpha_j z^j$ are outside of the unit circle. Bühlmann (1997, 1998), Kreiss (1988, 1992), and Paparoditis (1996) proposed approximating (7.2) by an AR($p$) model in which $p = p(n)$ increases with increasing sample size $n$. Let $\{a_{nj} : j = 1,...,p\}$ denote least squares or Yule-Walker estimates of the coefficients of the approximating process, and let $\{\hat{U}_{nj}\}$ denote the centered residuals. This procedure, which is called the sieve bootstrap, consists of generating bootstrap samples according to the process

$$(8.3) \quad \hat{X}_i - m = \sum_{j=1}^{p} a_{nj}(\hat{X}_{i-j} - m) + \hat{U}_j,$$

where $m = n^{-1}\sum_{i=1}^{n} X_i$ and the $\hat{U}_j$'s are sampled randomly with replacement from the $U_{nj}$'s. Bühlmann (1997), Kreiss (1992, 2000), and Paparoditis (1996) give conditions under which this procedure consistently estimates the distributions of sample averages, sample autocovariances and autocorrelations, and the regression coefficients $a_{nj}$ among other statistics.

Choi and Hall (2000) investigated the ability of the sieve bootstrap to provide asymptotic refinements to the coverage probability of a one-sided confidence interval for the mean of a linear statistic when the $X_i$'s are scalar random variables. A linear statistic has the form

$$\theta_n = (n - q + 1)^{-1} \sum_{i=1}^{n-q+1} G(X_i, ..., X_{i+q-1}),$$

where $q \geq 1$ is a fixed integer and $G$ is a known function. Define $\theta = E[G(X_1,...,X_q)]$. Choi and Hall (2000) considered the problem of finding a one-sided confidence interval for $\theta$. They gave conditions under which the difference between the true an nominal coverage probability of this interval is $O(n^{-1+\varepsilon})$ for any $\varepsilon > 0$. This is only slightly larger than the difference of $O(n^{-1})$ that is available with independent data.

A third possibility is that the DGP is a Markov process or can be approximated by such a process. The class of Markov and approximate Markov processes contains ARCH, GARCH, and many other processes that are important in applications. The assumption that the DGP is a Markov or approximate Markov process is weaker and, therefore, more general than the assumption that the DGP belongs to a finite-dimensional parametric family or is linear.

When the DGP is a stationary Markov process, the bootstrap can be implemented by estimating the Markov transition density nonparametrically. Bootstrap samples are generated by the stochastic process implied by the estimated transition density. Call this procedure the Markov bootstrap (MB). The MB was proposed by Rajarshi (1990), who gave conditions under which it consistently estimates the asymptotic distribution of a statistic. Datta and McCormick (1995) gave conditions under which the error in the MB estimator of the distribution function of a normalized sample average is almost surely $o(n^{-1/2})$.

Horowitz (2003) investigates the ability of the MB to provide asymptotic refinements for confidence intervals and tests based on Studentized statistics of the form

$$t_n = T^{1/2}[H(m) - H(\mu)] / s_T,$$

where $H$ is a smooth function, $\mu = E(X_1)$, $m = T^{-1} \sum_{t=1}^{T} X_t$, and $s_T^2$ is a consistent estimator of the variance of the asymptotic distribution of $T^{1/2}[H(m) - H(\mu)]$. Thus, for example, a symmetrical $1-\alpha$ confidence interval for $H(\mu)$ is

$$H(m) - z_{1-\alpha} s_T \leq H(\mu) \leq H(m) + z_{1-\alpha} s_T,$$

where $z_{1-\alpha}$ is a critical value. Horowitz (2003) gives conditions under which the difference between the true and nominal coverage probabilities of a symmetrical confidence interval is $O(n^{-3/2+\varepsilon})$ for any $\varepsilon > 0$. The difference for a one-sided confidence interval is $O(n^{-1+\varepsilon})$. In contrast, the asymptotic normal approximation makes errors of size $O(n^{-1})$ and $O(n^{-1/2})$, respectively, for symmetrical and one-sided confidence intervals.

The block bootstrap is a method that makes even weaker assumptions about the DGP. This method consists of dividing the data into blocks and sampling the blocks randomly with replacement. The blocks may be non-overlapping (Carlstein 1986, Hall 1985) or overlapping (Hall 1985, Künsch 1989, Politis and Romano 1993). Andrews (2004) and Politis and Romano (1993) describe variants of these blocking schemes. Details of the various blocking schemes are presented in the foregoing references and in Härdle, Horowitz, and Kreiss (2003); Hall and Horowitz (1996); and Lahiri (2003). The root mean-square errors of bootstrap estimators of distribution functions are smaller with overlapping blocks than with non-overlapping ones. This suggests that overlapping blocks are preferred for applications, although the differences between the numerical results obtained with the two types of blocking are often very small (Andrews 2004). The rates of convergence of the errors made with overlapping and non-overlapping blocks are the same.

Regardless of the blocking method that is used, the block length must increase with increasing sample size $n$ to make bootstrap estimators of moments and distribution functions consistent. The block length that minimizes mean square estimation error depends on what is being estimated. Hall, Horowitz, and Jing (1995) showed that with either overlapping or non-overlapping blocks, the optimal block-length is $l \propto n^r$, where $r = 1/3$ for estimating bias or variance, $r = 1/4$ for estimating a one-sided distribution function (e.g., $P(t_n \leq \tau)$), and $r = 1/5$ for estimating a symmetrical distribution function (e.g., $P(|t_n| \leq \tau)$). The results obtained with the block bootstrap can be sensitive to the choice of block length. Hall, Horowitz, and Jing (1995) and Lahiri (2003) describe data-based methods for choosing the block length in applications.

Block bootstrap sampling does not exactly replicate the dependence structure of the original data-generation process. For example, if nonoverlapping blocks are used, bootstrap observations that belong to the same block are deterministically related, whereas observations that belong to different blocks are independent. This dependence structure is unlikely to be present in the original data-generation process. As a result, the finite-sample covariance matrices of the asymptotic forms of parameter estimators obtained from the original sample and from the bootstrap sample are different. The practical consequence of this difference is that asymptotic refinements cannot be obtained by applying the usual formulae for test statistics to the block-bootstrap sample. It is necessary to develop special formulae for the bootstrap versions of test statistics. These formulae contain factors that correct for the differences between the asymptotic covariances of the original-sample and bootstrap versions of test statistics without distorting the higher-order terms of asymptotic expansions that produce refinements. Härdle, Horowitz, and Kreiss (2003); Hall and Horowitz (1996); and Lahiri (2003), among other references cited in this section, describe the appropriate versions of test statistics and explain why they are needed.

The asymptotic refinements provided by the block bootstrap depend in a complicated way on what is being estimated. Andrews (2004), Härdle, Horowitz, and Kreiss (2003); Hall and Horowitz (1996); and Hall, Horowitz, and Jing (1995), among other references cited in this section, provide details and numerical illustrations. The estimation errors made by the block bootstrap converge more slowly than the errors made by methods that make stronger assumptions about the DGP but more rapidly than the errors made by asymptotic normal approximations.

## 9. CONCLUSIONS

The bootstrap consistently estimates the asymptotic distributions of econometric estimators and test statistics under conditions that are sufficiently general to accommodate most applications. To achieve consistency, however, the bootstrap sampling procedure must be matched to the application. Often, it suffices to sample one's data randomly with replacement, but there are important cases that require sampling procedures that are more complex and/or modifications of the estimation procedure. This article has reviewed several examples of such cases. When used properly, the bootstrap provides a way to substitute computation for mathematical analysis if analytical calculation of the asymptotic distribution of an estimator or test statistic is difficult or impossible. The bootstrap is more accurate than subsampling in settings where the bootstrap is consistent and the relative accuracy of the two methods is known.

Under conditions that are stronger than those required for consistency but still general enough to accommodate many econometric applications, the bootstrap provides a more accurate approximation to the finite-sample distribution of an estimator or test statistic than does first-order asymptotic theory. The approximations of first-order asymptotic theory are often inaccurate with samples of the sizes encountered in applications. As a result, the errors in the rejection probabilities of hypothesis tests and the coverage probabilities of confidence intervals based on first-order approximations can be very large. The bootstrap can provide dramatic reductions in these errors. In many cases of practical importance, the bootstrap essentially eliminates finite-sample errors in rejection and coverage probabilities.

This article has emphasized the need for care in applying the bootstrap and the importance of asymptotically pivotal statistic for obtaining asymptotic refinements. However, even if asymptotic refinements are not available or not desired, the bootstrap should be used to obtain confidence intervals and critical values for hypothesis tests, not standard errors. Standard errors are important because they bear a simple relation to quantiles of the normal distribution, but bootstrap distributional approximations are non-normal. There is no simple relation between standard errors and quantiles of bootstrap distributions.

Donald, S.G. and H.J. Paarsch (1996). Identification, estimation, and testing in empirical models of auctions within the independent private values paradigm. *Econometric Theory*, 12, 517-567.

Efron, B. (1979), Bootstrap methods: another look at the jackknife. *Annals of Statistics* 7:1-26.

Efron, B. (1987), Better bootstrap confidence intervals. *Journal of the American Statistical Association*, 82:171-185.

Efron, B. and R.J. Tibshirani (1993), *An Introduction to the Bootstrap*. Chapman & Hall, New York.

Eicker, F. (1963). Asymptotic normality and consistency of the least squares estimators for families of linear regressions. *Annals of Mathematical Statistics*, 34, 447-456.

Eicker, F. (1967). Limit theorems for regression with unequal and dependent errors. In L. LeCam and J. Neyman (eds.), *Proceedings of the 5th Berkeley Symposium on Mathematical Statistics and Probability*, pp. 59-82. University of California Press, Berkeley, CA.

Fan, J. and I. Gijbels (1996). *Local Polynomial Modelling and Its Applications*. Chapman & Hall, New York.

Fan, J., T.-C. Hu, and Y.K. Truong (1994). Robust nonparametric function estimation. *Scandinavian Journal of Statistics*, 21, 433-446.

Flinn, C. and J. Heckman (1982). New methods for analyzing structural models of labor force dynamics. *Journal of Econometrics*, 18, 115-168.

Gill, R.D. (1989). Non- and semi-parametric maximum likelihood estimators and the von Mises method (Part 1). *Scandinavian Journal of Statistics*, 16, 97-128.

Härdle, W. (1990). *Applied Nonparametric Regression*. Cambridge University Press Cambridge, U.K.

Heckman, J.J, J. Smith, and N. Clements (1997). Making the most out of programme evaluations and social experiments: accounting for heterogeneity in programme impacts. *Review of Economic Studies*, 64, 487-535.

Hirano, K. and J.R. Porter (2003). Asymptotic efficiency in parametric structural models with parameter dependent support. *Econometrica*, 71, 1307-1338.

Hong, H. and J. Li (2015). The numerical bootstrap. Working paper, Department of Economics, Stanford University.

Horowitz, J.L. (1992). A smoothed maximum score estimator for the binary response model. *Econometrica*, 60, 505-531.

Horowitz, J.L. (1994). Bootstrap-based critical values for the information-matrix test. *Journal of Econometrics*, 61, 395-411.

Horowitz, J.L. (1997), "Bootstrap methods in econometrics: theory and numerical performance", in D.M. Kreps and K.F. Wallis, eds. *Advances in Economics and Econometrics: Theory and Applications*, Seventh World Congress, vol. 3, pp.188-222. Cambridge University Press, Cambridge, U.K.

Horowitz, J.L. (1998a). Bootstrap methods for covariance structures. *Journal of Human Resources*, 33, 39-61.

Horowitz, J.L. (1998b). Bootstrap methods for median regression models. *Econometrica*, 66, 1327-1351.

Horowitz, J.L. (2001). The Bootstrap. In J.L. Heckman and E. Leamer, eds., *Handbook of Econometrics*, vol. 5, pp. 3159-3228. Elsevier Science, B.V, Amsterdam.

Horowitz (2002). Bootstrap critical values for tests based on the smoothed maximum score estimator. *Journal of Econometrics*, 111, 141-167.
58